\documentclass[aps]{revtex4}

\usepackage{graphicx}

\begin{document}

% Use the \preprint command to place your local institutional report
% number in the upper righthand corner of the title page in preprint mode.
% Multiple \preprint commands are allowed.
% Use the 'preprintnumbers' class option to override journal defaults
% to display numbers if necessary
%\preprint{}

%Title of paper
\title{Nonlinear development of electron beam driven weak turbulence in an
  inhomogeneous plasma}

% repeat the \author .. \affiliation  etc. as needed
% \email, \thanks, \homepage, \altaffiliation all apply to the current
% author. Explanatory text should go in the []'s, actual e-mail
% address or url should go in the {}'s for \email and \homepage.
% Please use the appropriate macro foreach each type of information

% \affiliation command applies to all authors since the last
% \affiliation command. The \affiliation command should follow the
% other information
% \affiliation can be followed by \email, \homepage, \thanks as well.
\author{E.P.\ Kontar}
\email[]{eduardk@astro.uio.no}
%\homepage[]{Your web page}
%\thanks{}
%\altaffiliation{}
\affiliation{Institute of Theoretical Astrophysics, University of
Oslo, Postbox 1029, Blindern, 0315 Oslo, Norway}

\author{H.L.\ P{\'e}cseli}
\email[]{hanspe@aurora.uio.no}
%\homepage[]{Your web page}
%\thanks{}
%\altaffiliation{}
\affiliation{Plasma and Space Physics Section University of Oslo,
Postbox 1048, Blindern 0316 Oslo, Norway}

%Collaboration name if desired (requires use of superscriptaddress
%option in \documentclass). \noaffiliation is required (may also be
%used with the \author command).
%\collaboration can be followed by \email, \homepage, \thanks as well.
%\collaboration{}
%\noaffiliation

\date{\today}

\begin{abstract}
The self-consistent description of Langmuir wave and ion-sound
wave turbulence in the presence of an electron beam is presented
for inhomogeneous non-isothermal plasmas. Full numerical solutions
of the complete set of kinetic equations for electrons, Langmuir
waves, and ion-sound waves are obtained for a inhomogeneous
unmagnetized plasma. The result show that the presence of
inhomogeneity significantly changes the overall evolution of the
system. The inhomogeneity is effective in shifting the wavenumbers
of the Langmuir waves, and can thus switch between different
process governing the weakly turbulent state. The results can be applied to
a variety of plasma conditions, where we choose  solar coronal parameters as an
illustration, when performing the numerical analysis. 
\end{abstract}

% insert suggested PACS numbers in braces on next line
% insert suggested keywords - APS authors don't need to do this
%\keywords{}

%\maketitle must follow title, authors, abstract, \pacs, and \keywords
\maketitle

% body of paper here - Use proper section commands
% References should be done using the \cite, \ref, and \label commands

\section{Introduction}

Electrostatic electron plasma waves are easily excited in
controlled laboratory experiments \cite{ikezi,wong}, and in
naturally occurring plasmas, by for instance low density electron
beams \cite{fitze,lin,hospod_gurnett,kintner_et_al,stasiew}. The
evolution of weakly nonlinear coherent wave phenomena has been
extensively studied in simple geometries, and found to be well
described by standard analytical models \cite{pecseli,robinson}. Cases
where  electron plasma waves, or Langmuir waves, are excited in
plasmas made linearly unstable by  electron beams \cite{melrose}
are particularly interesting as being one of the first kinetic
instabilities where the nonlinear evolution described by quasi-linear  
theory  \cite{ved_vel,drum_pines} was studied experimentally in a 
controlled laboratory
experiment \cite{roberson}. This beam-excitation of plasma waves
has proved particularly interesting in space plasmas, where high
energy, low density electron bursts or beams are abundant. 
Full-wave numerical solutions for such problems have been carried out
in unmagnetized as well as magnetized \cite{hafizi,jgr_goldman} plasmas, 
based on a standard and widely accepted theoretical model \cite{zakharov}.
It is
interesting that beam generated electron waves are observed
by satellites in the interplanetary plasma near 1 AU \cite{lin},
and in the ionosphere at high as well as lower altitudes by
sounding rockets as well as satellites
\cite{lin,hospod_gurnett,kintner_et_al,ergun,bonell}. In these and
similar cases, the waves appear as irregular or bursty in nature.
A description aiming at a general full wave analysis seems
unrealistic in such cases, and a weak turbulence model has been
suggested to give the evolution of the most important quantities,
such as the averaged electron distribution function, and the wave
spectrum \cite{Tsytovich,camac,Vedenov67}. The beam-plasma
instability excites a spectrum of Langmuir waves, which in turn
are damped or converted via nonlinear plasma processes, such as
scattering off ions, $l+i\rightarrow l^{\prime}+i^{\prime}$
(nonlinear Landau damping off ions \cite{Tsytovich,dysthe}), and
decay into a Langmuir and ion sound waves, $l\rightarrow l+s$,
discussed previously \cite{Tsytovich,zakh_rep}. Obviously, there
are also other nonlinear processes that change the spectrum of
Langmuir waves, but the processes mentioned are much more
effective in affecting Langmuir waves in the applications of
interest.

In the present work, we consider the case where energy density of
the electron beam is much smaller than the thermal energy density
of background plasma. The plasma waves and electron beam are
described self-consistently by weak turbulence theory
\cite{Tsytovich}. Quasi-linear relaxation of an electron beam with
the velocity $v_b>>v_{Te}$ generate the primary Langmuir waves
with wavenumbers ${\bf k}\approx \omega _{pe}{\bf v}_b/v_b^2$ that
causes electron beam electron distribution function to relax
toward a plateau in velocity space ranging from $v\sim v_b$ down
to $v\sim v_{Te}$. Nonlinear processes, $l+i\rightarrow
l^{\prime}+i^{\prime}$ and $l\rightarrow l^{\prime}+s$, are
effective in scattering of primary, beam generated waves with
wavenumber ${\bf k}$, into secondary Langmuir waves with
wavenumber ${\bf k}^{\prime}\approx {\bf -k}$. The decay of a
Langmuir wave also leads generation of the ion-sound waves with
${\bf k}_s\approx 2{\bf k}$. The process repeats and produce the
next generation of Langmuir waves. Every elementary cascade
decreases the absolute value of Langmuir wave number by a small
value $k^{*}_d=2\sqrt{m_e/m_i}\sqrt{1+3T_i/T_e}/(3\lambda_{De})$
for decay and
$k^{*}_s=2\sqrt{m_e/m_i}\sqrt{T_i/T_e}/(3\lambda_{De})$ for
scattering. Therefore repeated scattering and decay processes lead
to Langmuir waves being accumulated in the region of small values
of $k\leq k^{*}$, the  so-called Langmuir wave condensate. The
decay processes are considered to be dominant process for
non-isothermal plasma $T_e>>T_i$, whereas scattering off ions is
more important for isothermal plasma where ion-sound waves are
heavily damped.

However, the presence of plasma density gradient can change the
spectrum of Langmuir waves significantly. In the presence of a
plasma gradient, a given Langmuir wave with wavenumber ${\bf k}$
propagating in inhomogeneous plasma changes its wave number to
${\bf k}\pm \Delta{\bf k}$, where $\Delta {\bf k}$ is determined
by a plasma gradient. This effect may significantly change the
Langmuir wave spectrum, and therefore to slow down the nonlinear
processes. If the plasma gradient is opposite to the direction of
beam propagation, and the drift rate due to inhomogeneity is
comparable with rate of nonlinear processes, a steady state
spectrum of Langmuir waves can appear. This case was considered in
the case of isothermal plasma when the scattering off ions is the
only nonlinear process in \cite{Sakharov79}. It was shown that the
condensate of Langmuir waves appears, but in a shifted region of
$k$ and scattering off ions can be compensated by a plasma
inhomogeneity. Moreover, a plasma inhomogeneity can change the
quasi-linear relaxation rate by shifting plasma waves from the
unstable phase velocity region  to the region where Langmuir waves
are strongly damped by the beam electrons \cite{Ryutov69}. In a
number of studies \cite{Breizman69} it was found that a plasma
inhomogeneity may suppress quasi-linear relaxation, provided the
beam density is sufficiently low. The positive plasma gradient can
also lead to ``self-acceleration'' of beam electrons
\cite{Breizman69}. The importance of a plasma inhomogeneity
demonstrated in previous studies \cite{Kontar01b} and the presence
of inhomogeneities in most plasmas found in nature as well as in
laboratories stimulates an investigation of the time evolution of
beam-driven Langmuir turbulence in non-isothermal plasma. The principal aim
of the present paper is to discuss the effect of density gradients on the
wave dynamics and to
demonstrate that even weak gradients can be of profound importance.

In the present study we perform a self-consistent investigation of
the time evolution of weak turbulence in weakly inhomogeneous
plasma. It is shown that a plasma inhomogeneity significantly
modifies the scenario of weak turbulence. The disposition of the
paper is as follows. In Section~\ref{sec:prob} we present a
formulation of the problem, and present the basic equations. In
Section~\ref{sec:numsol} we present numerical results. In
Section~\ref{sec:discuss} we summarize the main results of our
investigations, while Section~\ref{sec:concl} contains our
conclusions. In the present paper we attempt to consider
physically realistic parameters, and have chosen those appropriate
for electron beams and plasmas for the conditions in the solar
corona. The problem is extremely time consuming 
numerically, and for this reason our investigations are restricted
to one spatial dimension.

\section{Formulation of the problem}\label{sec:prob}

We consider a beam of fast electrons and plasma waves within the
limits of weak turbulence theory, when the energy density $W$ of
plasma waves with wave number $k$ is much less then that of
surrounding plasma
\begin{equation}\label{weak}
  \frac{W}{n\kappa T_e}<(k\lambda _{De})^2
\end{equation}
where $n$ and $T_e$, are the electron plasma density and temperature,
respectively,
$\lambda _{De}$ is the electron Debye length, while $\kappa$ is
Boltzmann's constant. In weak
turbulence theory the evolution of electrons and waves is
described by kinetic equations for an electron distribution
function and spectral energy densities of plasma waves. The
equations are essentially nonlinear, which significantly
complicates the problem \cite{Tsytovich}. However, having in mind
application to low-$\beta$ systems with relatively strong magnetic
field when the energy density of magnetic field is much larger
than the kinetic energy of fast electrons, but the magnetic field
is not strong enough to magnetize the plasma waves, we can treat
the system in one spatial dimension. The electron beams in the
solar corona plasma are typical examples for such systems
\cite{Melrose90}. In these and similar cases, the plasma
inhomogeneity along the beam propagation can not be ignored.

We also assume that the variation $\mbox{d}\lambda$ of the wave
length $\lambda$ of the Langmuir  oscillations is a small, i.e.\
\begin{equation}\label{WKB}
\left|\frac{\mbox{d}\lambda}{\mbox{d} x}\right|\ll 1
\end{equation}
or in other words, we describe wave propagation in geometrical
optics (WKB) approximation \cite{Vedenov67,Galeev63}. Using the
fact that the frequency of a Langmuir wave does not change during
its propagation in the plasma, we readily derive the condition for
applicability of the WKB approximation from (\ref{WKB})
\begin{equation}\label{cond}
  \frac{v}{|L|} \ll 3\omega_{pe}(x)\left(\frac{v_{Te}}{v}\right)^2
\end{equation}
where
\begin{equation}\label{L}
L\equiv \omega_{pe}(x)\left(\frac{\partial \omega
_{pe}(x)}{\partial x} \right)^{-1}
\end{equation}
is the scale of the local inhomogeneity, $\omega _{pe}$ is the
local plasma frequency, and $v_{Te}$, $v=\omega_{pe}/k$ are the
electron thermal and wave phase velocities, respectively.

The time evolution of the average velocity distribution $f(v,t)$
is described by quasi-linear theory, which basically describes a
diffusion process in velocity space, where the diffusion
coefficient is self-consistently determined by the spectrum of the
Langmuir waves \cite{ved_vel,drum_pines}. The velocity
distribution is assumed to be the same all over the relevant part
of space, and in those cases where we deal with an inhomogeneous plasma,
the density variation is contained in a coefficient. The growth
and damping of plasma waves is accounted for by the standard
Landau prescription and derived from the derivative of the
velocity distribution at the phase velocity of the waves. The
evolution of the wave spectra is described by the wave kinetic
equation basically having the form
\begin{equation}
\frac{\partial W_k}{\partial t} +v_g \frac{\partial W_k}{\partial
x} - \frac{\partial \omega_k}{\partial x}\frac{\partial
W_k}{\partial k}=St \, ,\label{wave_kin}
\end{equation}
which for $St=0$ is the Liouville equation. Equation (\ref{wave_kin}) can be
postulated from basic physical arguments \cite{camac,Vedenov67},
or be derived \cite{hans} from more basic equations
\cite{zakharov}. In the dynamic equation (\ref{wave_kin}), we can
heuristically interpret $W_k$ as a space time varying distribution
of wavepackets, each with a carrier wavenumber $k$, propagating
with a corresponding group velocity $v_g\equiv {\partial
\omega_k}/{\partial k}$, and subject to an effective force
$-{\partial \omega_k}/{\partial x}$. In the present case we have
${\partial \omega_k}/{\partial x}\approx {\partial
\omega_{pe}}/{\partial x}$, and this  force is directly related to
the gradient in plasma density through the density dependence of
$\omega_{pe}$. The wavepackets constituting $W_k$ are accelerated
in the direction towards smaller densities while the carrier
wavenumber increases. The term $St$ on the right hand side of
(\ref{wave_kin}) accounts for sources and sinks, together with
effects that redistribute energy within the spectrum. With the
foregoing assumptions (\ref{WKB}), we can assume ${\partial
\omega_{pe}}/{\partial x}\approx \omega_{pe}/L$, where
$\omega_{pe}$ is now a constant plasma frequency obtained at a
reference position in the center of the system. Apart from a
negligible correction of the relative order $k\lambda_{De}$, this
effective force is independent of wavenumber, and it will not
induce any spatial variations in an initially uniform distribution
$W_k$. For that case we consequently have ${\partial
W_k}/{\partial x}=0$. With these assumptions we simplify
(\ref{wave_kin}) in the following. As far as $St$ is concerned, we
have the linear instability acting as a source, and Landau damping
as a sink of wave energy. Decay and nonlinear Landau damping act
as sinks of wave energy, but equally important, these effects
redistribute the energy within the spectrum $W_k$. The
mathematical expressions for these latter effects are standard
\cite{Tsytovich,zakh_rep}.

The basic equations are treated as an initial 
value problem (just as in related recent studies 
\cite{forme,cairns_00,yoon_01}). This gives a simplified alternative to the
full problem, retaining at the same time the physics being important here. 
Using the assumptions mentioned above we can write the system of
kinetic equations of weak turbulence theory

\begin{equation}
\frac{\partial f}{\partial t}= \frac{4\pi^2 e^2
}{m^2}\frac{\partial}{\partial v} \frac{W_k}{v}\frac{\partial
f}{\partial v},\;\;\; \label{eqk1}
\end{equation}
\begin{equation}
\frac{\partial W_k}{\partial t}-\frac{\omega_{pe}}{L}
\frac{\partial W_k}{\partial k} =\frac{\pi
\omega_{pe}^3}{nk^2}W_k\frac{\partial f}{\partial
v}+St_{ion}(W_k)+St_{decay}(W_k,W^s_k) \label{eqk2}
\end{equation}
\begin{eqnarray}
\frac{\partial W^{s}_{k}}{\partial t}=-2\gamma^s_k
W^{s}_{k}-\alpha {\omega^s_k}^2\int\left(
\frac{W_{k^{\prime}-k}}{\omega_{k^{\prime}-k}}\frac{W^s_{k}}{\omega^s_{k}}-
\frac{W_{k^{\prime}}}{\omega_{k^{\prime}}}\left(\frac{W_{k^{\prime}-k}}{\omega_{k^{\prime}-k}}+
\frac{W^s_{k}}{\omega^s_{k}}\right)\right)\delta
(\omega_{k^{\prime}}-\omega_{k^{\prime}-k}-\omega^s_k)dk^{\prime}
\end{eqnarray}
\begin{eqnarray}
St_{decay}(W_k,W^s_k)=\alpha\omega_{k}
\int\omega^s_{k^{\prime}}\left[ \left(
\frac{W_{k-k^{\prime}}}{\omega_{k-k^{\prime}}}\frac{W^s_{k'}}{\omega^s_{k'}}-
\frac{W_k}{\omega_k}\left(\frac{W_{k-k^{\prime}}}{\omega_{k-k^{\prime}}}+
\frac{W^s_{k^{\prime}}}{\omega^s_{k^{\prime}}}\right)\right)\delta
(\omega_{k}-\omega_{k-k^{\prime}}-\omega^s_{k^{\prime}})-\right.\cr
\left.\left(
\frac{W_{k+k^{\prime}}}{\omega_{k+k^{\prime}}}\frac{W^s_{k'}}{\omega^s_{k'}}-
\frac{W_k}{\omega_k}\left(\frac{W_{k+k^{\prime}}}{\omega_{k+k^{\prime}}}-
\frac{W^s_{k^{\prime}}}{\omega^s_{k^{\prime}}}\right)\right)\delta
(\omega_{k}-\omega_{k+k^{\prime}}+\omega^s_{k^{\prime}})\right]
dk^{\prime}
\end{eqnarray}
\begin{eqnarray}
St_{ion}(W_k)=\beta \omega_k\int
\frac{(\omega_{k^{\prime}}-\omega_k)}{v_{Ti}|k-k^{\prime}|}
\frac{W_{k^{\prime}}}{\omega_{k^{\prime}}} \frac{W_k} {\omega_k}
\exp\left[-\frac{(\omega_{k^{\prime}}-\omega_k)^2}{2v_{Ti}^2|k-k^{\prime}|^2}\right]
dk^{\prime}
\end{eqnarray}
where
\begin{equation}
\alpha=\frac{\pi \omega^2_{pe}(1+3T_i/T_e)}{4n\kappa T_e},\;\;\;
\beta=\frac{\sqrt{2\pi}\omega^2_{pe}}{4n\kappa T_i(1+T_e/T_i)^2},
\end{equation}
\begin{equation}\label{gam_sk}
\gamma^s_k=\sqrt{\frac{\pi}{8}}\omega^s_k\left[\frac{v_s}{v_{Te}}+\left(\frac{\omega
^s_k}{kv_{Ti}}\right)^3\exp\left[- \left(\frac{ \omega ^s_k}{
kv_{Ti} }\right)^2 \right]\right]
\end{equation}
where $\gamma^s_k$ is the damping rate of ion-sound waves,
$v_s=\sqrt{\kappa T_e(1+3T_i/T_e)/m_i}$ is sound speed, $f(v,t)$ is the
averaged electron distribution function, $W(k,t)$ and $W^s(k,t)$
are the spectral energy densities of Langmuir waves and ion-sound
waves respectively. $W(k,t)$ plays the same role for waves as the
electron distribution function for particles. The system
(\ref{eqk1}) and (\ref{eqk2}) describes the resonant interaction
$\omega_{pe}=kv$ of electrons and Langmuir waves. The last term on
the left-hand side of (\ref{eqk2}) represent the  shift in wave
number induced by the plasma inhomogeneity.

In the right-hand side of equations (\ref{eqk1}) and (\ref{eqk2})
we omitted terms accounting for spontaneous emissions,  because
they are small  in comparison to the ones retained. It should be
also noted that small collisional terms are not included either.
The system reaches steady state before collisional effects become
important.

\section{Numerical results}\label{sec:numsol}

We consider an initial value problem when all initial energy is
accumulated in electron beam. The electron distribution function
of electrons at the initial time moment $t=0$ is the combination
of fast electrons and background Maxwellian electrons
\begin{equation}\label{eq_f0}
  f(v,t=0) = \frac{n_b}{\sqrt{\pi}\Delta
  v_b}\exp\left(-\frac{(v-v_b)^2}{\Delta
  v_b^2}\right)+ \frac{n}{\sqrt{2\pi}
  v_{Te}}\exp\left(-\frac{mv^2}{2\kappa T_e}\right)
\end{equation}
where  $n_b$, $\Delta v_b$ are the beam density and the electron
beam thermal velocity.  The initial spectral energy density is
thermal
\begin{equation}
W(k,t=0)\approx \frac{\kappa T_e}{2\pi ^2 \lambda _{De}^2}, \label{eq:5}
\end{equation}
where $T_e$ is the electron temperature of the surrounding plasma,
and $\lambda _{De}$ is the electron Debye length. The system of
kinetic equations is reduced to dimensionless form and integrated
using finite difference schemes. For the numerical time
integration we used a method similar to the one used in
\cite{Payne89} and the quasi-linear terms are integrated using
methods described in \cite{Kontar01}. Previous results
\cite{Payne89,Kontar01} are found in full agreement with the
corresponding limiting cases of our calculations.

The electron beam and plasma parameters are taken typical for the
conditions in the solar corona \cite{Kontar_sun,sol_phys}.
We use $v_b=12.8\,v_{Te}=5\times 10^9$ cm/s,
$\Delta v_b=0.3v_b$, $T_i/T_e=0.3$, $T_e=10^6$ K, $n_b=50$
cm$^{-3}$, and $\omega _{pe}/2\pi=200$ MHz. The plasma inhomogeneity
is given by a constant  plasma gradient, which can have either
sign. Six different cases have been considered: the strongest
plasma inhomogeneity considered is $L=\pm1\times 10^8$ cm, a
medium plasma inhomogeneity $L=\pm 5\times 10^8$ cm, and a weak
plasma inhomogeneity $L=\pm1\times 10^9$ cm. 
Such plasma inhomogeneity might exist in the low corona \cite{Benz},
in ionosphere \cite{ISCAT}, and in the solar corona due to small scale 
inhomogeneity \cite{Smith79}.

\subsection{Homogeneous plasma}

\begin{figure}
\begin{center}
\includegraphics[width=78mm]{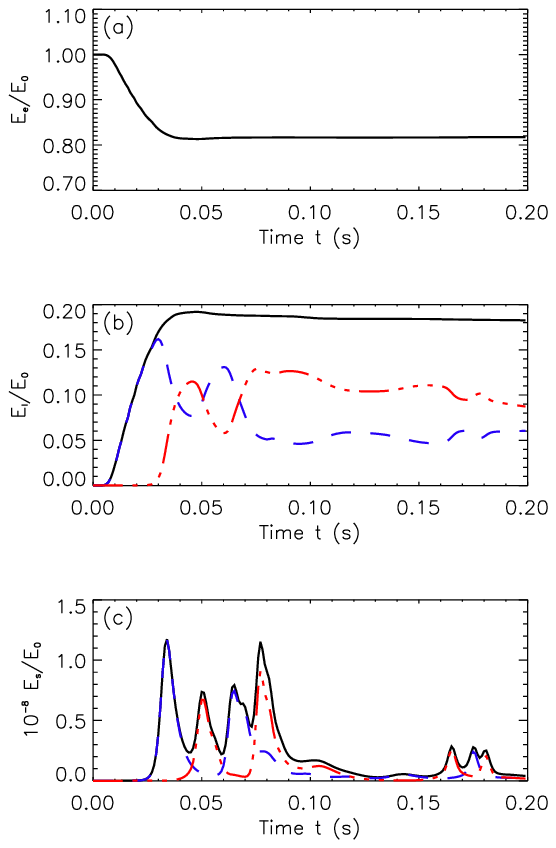}
\includegraphics[width=78mm]{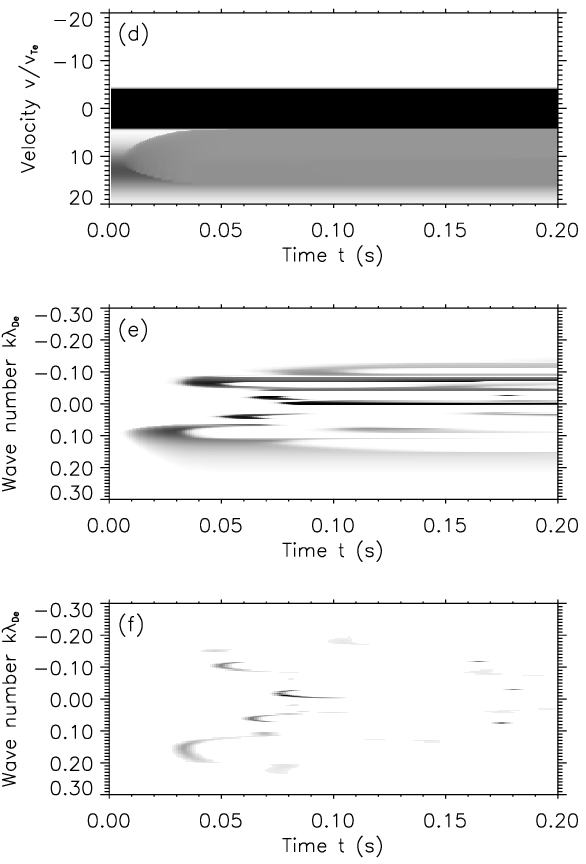}
\end{center}
\caption{Homogeneous density plasma. The normalized energy of electrons
$E_e(t)/E_0$, is shown in (a), the normalized energy in the Langmuir waves
$E_l(t)/E_0$ in (b), and correspondingly for the ion-sound waves
$E_s(t)/E_0$ are shown in (c). The energy of
waves along $E_{l,s}^{+}/E_0$  and against
$E_{l,s}^{+}/E_0$ the beam propagation is given by  dashed  and
dash-dotted lines, respectively. The time
evolution of the spectral distribution of electrons
$f(v,t)v_b\sqrt{\pi}/n_b$ is shown in (d), for Langmuir waves
$W(k,t)\omega_{pe}/mn_bv_b^3$ in (e), and ion-sound waves
$W(k,t)\omega_{pe}/mn_bv_b^3$ is shown in (f). The disposition of the
following figures is the same as here.} \label{fig1}
\end{figure}

As a reference case we first considered the evolution of Langmuir
turbulence in a homogeneous plasma, corresponding to $L\rightarrow
\infty$. Results are shown in fig.~\ref{fig1},  demonstrating that
the fastest process in the system is quasi-linear relaxation,
which drives energy out of a beam and into Langmuir waves. As a
result of quasi-linear relaxation the electron distribution
function rapidly flattens, building a plateau from
$15v_{Te}>v>4v_{Te}$ for $t\approx 0.02$ s, which is close to a
quasi-linear time $\tau \approx n/\omega _{pe}n_b$. The beam driven
Langmuir turbulence has a bright maximum at $k\lambda
_{De}=v_{Te}/v_b\approx 0.087$. The primary Langmuir waves
generated by a beam are subject to decay into a secondary Langmuir
waves and ion-sound waves. The decay of Langmuir waves starts from
the maximum of $W(k)$ and produce back-scattered Langmuir waves
with maximum at $k\lambda _{De}=-0.065$. Secondary Langmuir waves
reaches its maximum at $t=0.04$ s. In their turn secondary
Langmuir waves produce a new generation of scattered Langmuir
waves with maximum at $k\lambda _{De}\approx 0.044$. It is obvious
that an absolute value of $k$ decreases with each act of decay by
$k_d^{*}\lambda _{De}\approx 0.021$. Thus, step by step the
maximum of the Langmuir wave distribution approaches the region of
wavenumbers where decay is prohibited, i.e.\ the region
$k<k_d^*/2$. However, since the spectrum of Langmuir waves
generated by a beam is broad in wavenumbers, the decay continues
in wavenumber regions  where the level of Langmuir waves is
relatively low. The rate of three wave decay is proportional to
the intensity of Langmuir waves and therefore at a given moment
we see a few generations of Langmuir waves simultaneously
(fig.~\ref{fig1}). Thus, each generation of Langmuir waves appear
as a parabolic structure in that part of fig.~\ref{fig1} which shows the
wavenumber distribution as a function of time.

The ion-sound turbulence has a low intensity due to the strong ion
Landau damping in a plasma with $T_i/T_e=0.3$. However, ion-sound
waves are generated during each  decay. The decay of a primary
waves is marked by ion-sound waves with maximum at $k\lambda
_{De}\approx 0.15$, which corresponds to $k_s=2k-k_d^{*}$. The
maximum of ion sound waves due to the decay of secondary waves is
seen at $k_s\lambda_{De}\approx 0.125$.

The energy of fast electrons, i.e.\ the beam energy, is given by
\begin{equation}\label{ee}
E_e(t)=\int \limits _{4v_{Te}}^{\infty}mv^2f(v,t)dv/2 \, ,
\end{equation}
the total energy of waves
\begin{equation}\label{el}
E_{l,s}(t)= \int \limits _{-\infty}^{\infty}W^{l,s}(k,t)dk \, ,
\end{equation}
the energy of waves propagating along and against the beam
\begin{equation}\label{epm}
E^{+,-}_{l,s}(t)=\pm \int \limits _{0}^{\pm \infty}W^{l,s}(k,t)dk
\end{equation}
The corresponding distributions are shown in fig.~\ref{fig1}. The
lower integration limit in (\ref{ee}) of course chosen somewhat
arbitrarily, and the energy and density of a beam obtains a
somewhat arbitrary value. At later times,   $E_e$ may therefore
exceed its initial value, when some of the background electrons
are accelerated to velocities above the value $4v_{Te}$ chosen in
(\ref{ee}). 
In the presentation here,  the  distribution function  is truncated at a
level of 3 in the normalized units, and therefore the background plasma
appears as a black band. To interpret the grey-scale in, for instance, the
Langmuir wavenumber distribution as a function of time, the energy
distribution $E_l(t)/E_0$ can be used as guide.

The energy distribution in the system follows the well-known
scenario \cite{zakh_rep}. In the present case, one fifth of the
initial beam energy is transformed into wave energy, $E_e^+$, of
Langmuir waves along the beam propagation during the quasi-linear
relaxation. However, the decay redistributes the energy between
the primary waves and scattered waves. As a result the typical
oscillations of wave energy,  $E_l^{\pm}$, along and opposite to
the beam direction  are clearly seen in fig.~\ref{fig1}(b). The total
energy of Langmuir waves is almost constant, displaying the
conservation of energy in the three-wave decay. Due to the strong
damping of ion-sound waves, the energy of density oscillations is a
small fraction of the initial beam energy. Ion-sound waves
generated in the decay are rapidly absorbed by linear ion Landau
damping, which is significant for the temperature rations in the
present problem. Therefore the amplitude of ion-sound wave
oscillations is small, see fig.~\ref{fig1}(f). In order to interpret the
grey-scale intensity levels, the energy curves in figs.~\ref{fig1}(b)and
\ref{fig1}(c) can be used for an estimate.
Obviously, the system tends towards a steady state solution. 
The final energy distribution
of Langmuir waves is not symmetric in $k$. The energy of waves
propagating against beam direction, $E^{-}_l$, is one half of that
propagating along the beam, $E^{+}_l$. This is simply due to back
absorption of Langmuir waves by the beam.

\begin{figure}
\begin{center}
\includegraphics[width=78mm]{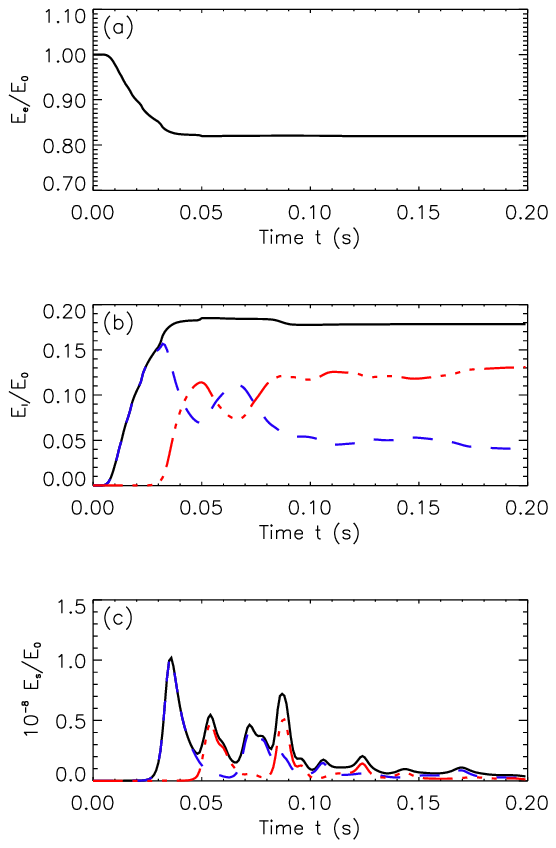}
\includegraphics[width=78mm]{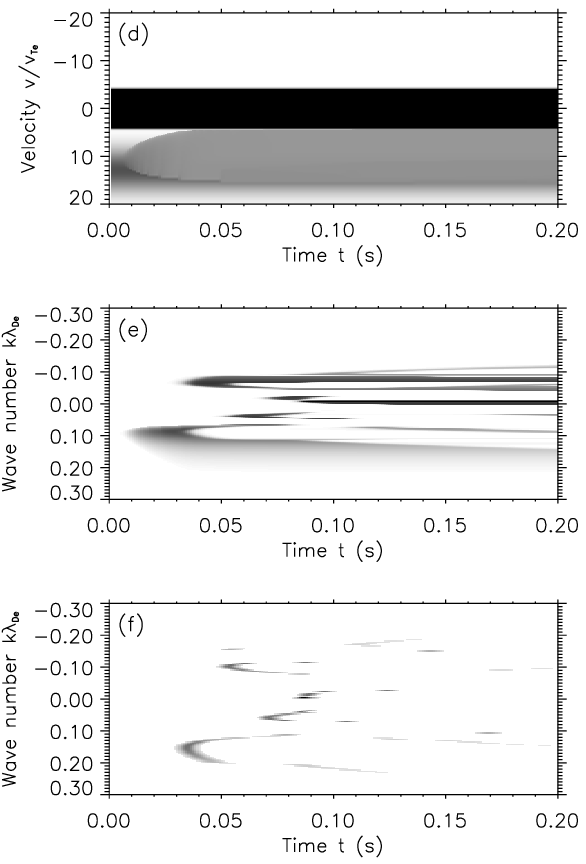}
\end{center}
\caption{The same as fig.~\ref{fig1}, but the scattering off ions is
switched off.} \label{fig2}
\end{figure}

The scattering off ions seems to be negligible during the initial
redistribution of Langmuir waves. This trivial result proves the
well-known fact that decay process is the fastest process changing
the Langmuir spectrum in non-isothermal plasma. However, the
scattering off ions plays a critical role at later stages, when
the Langmuir waves are accumulated in the region of $|k|\leq
k_d^*/2$. The reason for this is that decay effectively generates
high level of Langmuir waves in the region of small $k$ where
scattering off ions more effective. Moreover, since decay is
impossible for $k_s^*/2\leq |k|\leq k_d^*/2$, scattering off ions
seems to be the only process that continues to contribute to the
Langmuir wave condensate. To prove the role of scattering off
ions, we  turned off the corresponding term in our calculations.
These results are presented in fig.~\ref{fig2}. 
Some differences are found between figs.~\ref{fig1} and \ref{fig2},
e.g.\ details of the spectra of Langmuir waves presented for these two cases
differ for $t>0.05$ s.

\subsection{Strong plasma inhomogeneity}

Since overall evolution of the system strongly depends on the
plasma gradient value we consider three different length scales for the
inhomogeneity separately.
\begin{figure}
\begin{center}
\includegraphics[width=70mm]{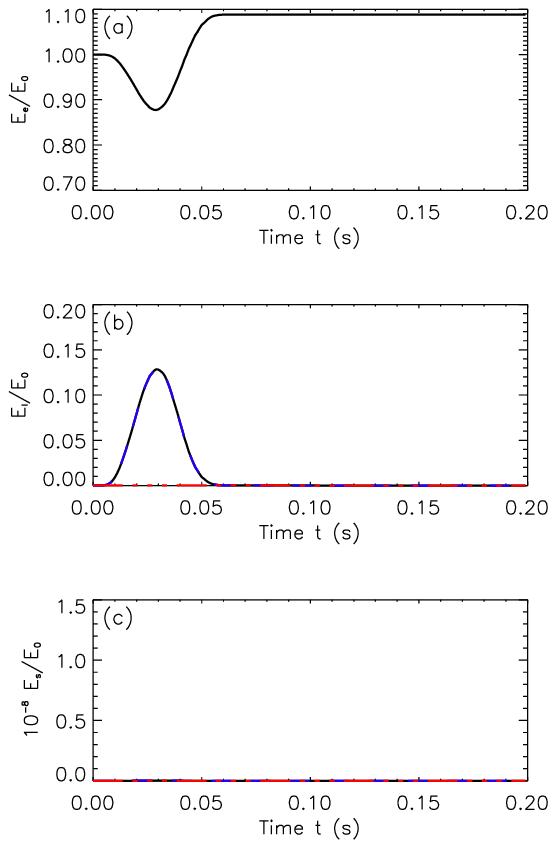}
\includegraphics[width=70mm]{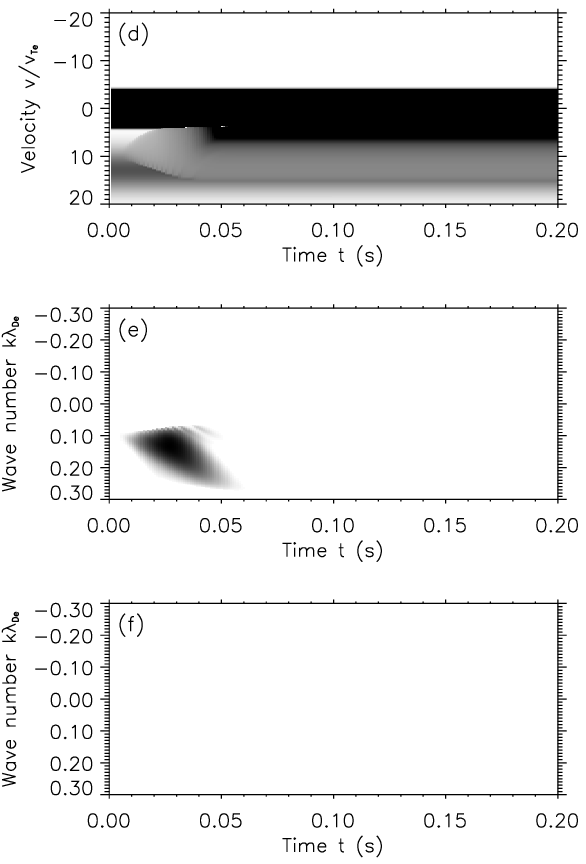}
\end{center}
\caption{The same as fig.~\ref{fig1}, but for decreasing density
plasma along beam propagation. Strong plasma inhomogeneity
$L=-1\times 10^8$ cm.}
 \label{fig3}
\end{figure}
\begin{figure}
\begin{center}
\includegraphics[width=70mm]{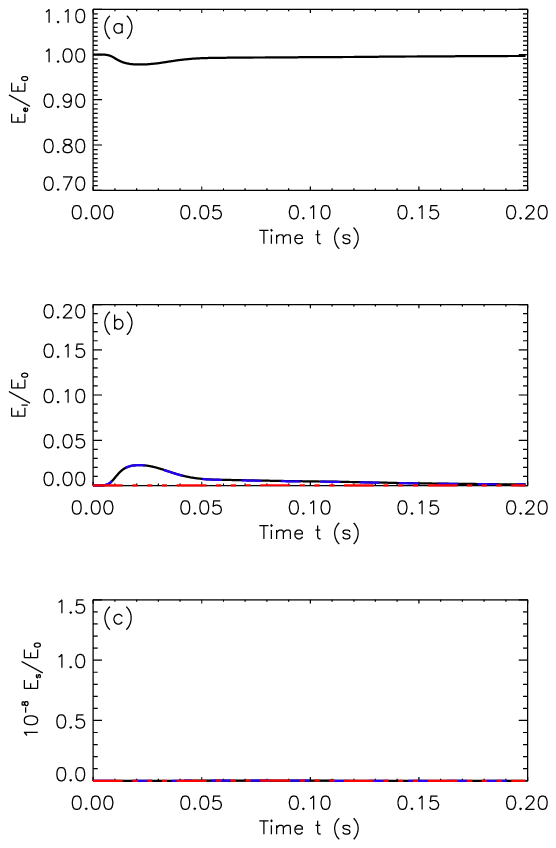}
\includegraphics[width=70mm]{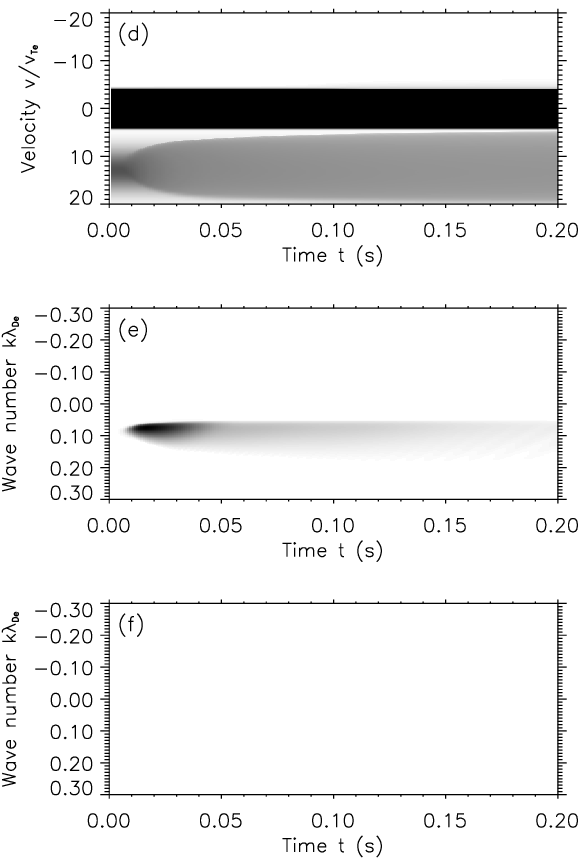}
\end{center}
 \caption{The same as fig.~\ref{fig1}, but for an increasing plasma density.
Strong plasma inhomogeneity, with $L=1\times 10^8$ cm.}
\label{fig4}
\end{figure}

The main results for strong plasma gradient are presented in
figs.~\ref{fig3} and \ref{fig4}. In this case the shift of the
spectrum, due to plasma inhomogeneity, is stronger than any
nonlinear process in the system, and the plasma inhomogeneity
suppresses all nonlinear processes, but it is not strong enough to
arrest the quasi-linear relaxation. Independent on the sign of the
plasma gradient, $L$, the evolution of the Langmuir turbulence is very
limited in time,  while the physical processes are different for
positive and negative gradients.

In the case of a positive plasma density gradient, see fig.~\ref{fig4}, 
all plasma waves generated during the relaxation are
absorbed back by the beam at time $t=0.02$ s. We see that the
drift of Langmuir waves in $k$ space is so fast that nonlinear
processes are not observable. No signs of ion-sounds waves are
observed neither. For the time comparable with the quasi-linear
time, Langmuir waves are shifted from the generation region to
absorption region. As a result, accelerated electrons are seen in
fig.~\ref{fig4}. The plateau is now formed from $20 \,
v_{Te}>v>4\,v_{Te}$. The amount of energy released from a beam is
a small fraction of the beam energy. Therefore, this case is
specially interesting in terms of stabilization of quasi-linear
relaxation by a plasma inhomogeneity. This limiting case has been
considered in the literature \cite{Ryutov69,Breizman69,Krasovskii78}
and a simplified solution can be found in this case. Thus, the
strong positive gradient $L=1\times 10^8$ cm leads to
self-acceleration of electrons in a beam, consistent with 
observations reported in \cite{lin}, for instance.

In the case of a negative plasma gradient (decreasing plasma
density), Langmuir waves are also effectively absorbed back by
electrons. Contrary to the case with $L>0$, Langmuir waves are now
absorbed by the background electrons with velocities close to thermal
velocities (classical Landau damping). As a result of this Landau
damping, we see the appearance of  accelerated electrons in the
range $4\,v_{Te}<v<9\,v_{Te}$. The electron distribution function
of these electrons is a decreasing function of velocity. Thus, due
to the inhomogeneity, plasma waves play a role of Dreicer  field
extracting electrons from the background Maxwellian distribution.
Figure~\ref{fig4} demonstrates that the amount of energy of
electrons with $v>4\,v_{Te}$ is greater than it was before the
quasi-linear relaxation. The spectral energy density of Langmuir
waves reaches its maximum value at $k\lambda _{De}\approx 0.15$,
which is quite different from the beam resonant region of $k$,
where generation of Langmuir waves takes place, see fig.~\ref{fig5}.
It is also worth noting that the amount of energy pumped into
Langmuir waves is much larger than in the case of a positive plasma
gradient and comparable with the case of a homogeneous plasma. The
amount of energy obtained by Langmuir waves is about 13\% of
initial beam energy. The maximum is reached at time 
$t=0.03$ s. Similar to the case of a positive plasma gradient, there
is no sign of either back-scattered Langmuir nor ion-sound waves.
Thus, the plasma inhomogeneity suppresses any further nonlinear
development of weak turbulence.

\subsection{Medium plasma inhomogeneity}

\begin{figure}
\begin{center}
\includegraphics[width=70mm]{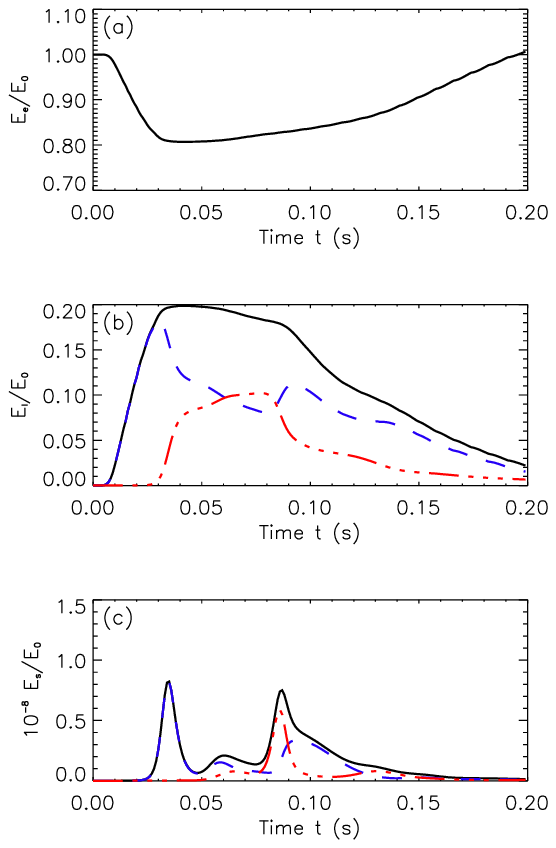}
\includegraphics[width=70mm]{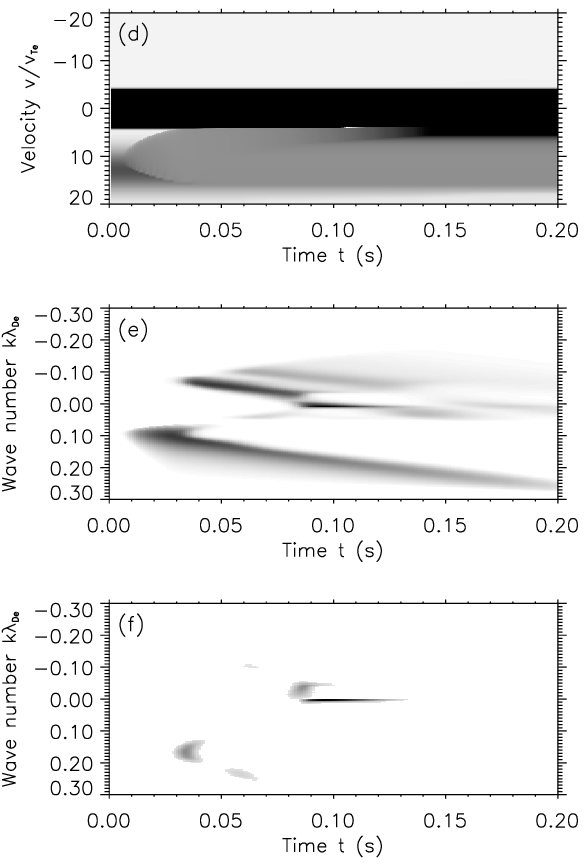}
\end{center}
 \caption{The same as fig.~\ref{fig1} but for a decreasing  plasma density.
Medium plasma inhomogeneity $L=-5\times 10^8$ cm.}
 \label{fig5}
\end{figure}

\begin{figure}
\begin{center}
\includegraphics[width=70mm]{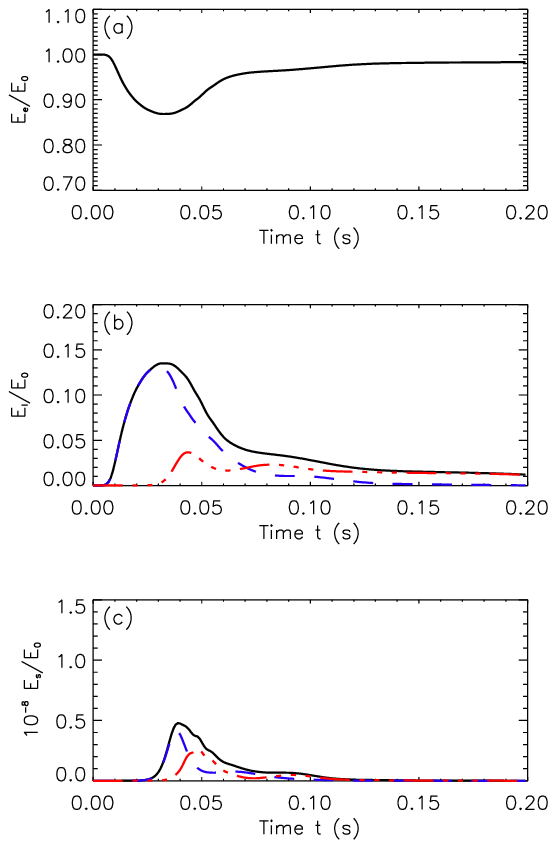}
\includegraphics[width=70mm]{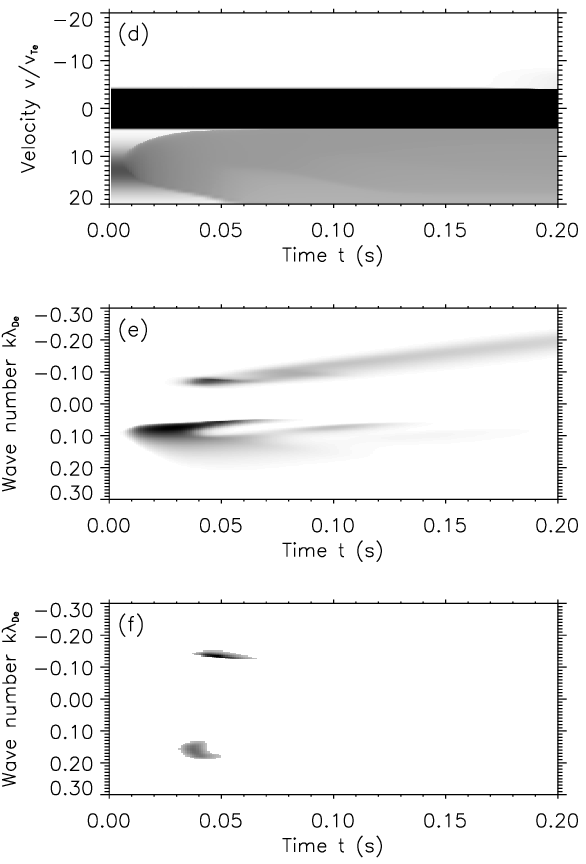}
\end{center}
 \caption{The same as fig.~\ref{fig1} but for an increasing plasma density.
Medium plasma inhomogeneity $L=5\times 10^8$ cm.}
 \label{fig6}
\end{figure}

Decreasing the value of the plasma gradient in our system we can
stimulate a further development of weak turbulence. For $|L| =
5\times 10^8$ cm a single decay is observable in figs.~\ref{fig5}
and \ref{fig6}. The primary waves, as in the case of homogeneous
plasma, have their maximum close to $k\lambda _{De}=0.086$. The
secondary waves are accumulated near $k\lambda _{De}=-0.065$. The
shift of the Langmuir wave spectrum due to plasma inhomogeneity is
comparable with three wave decay of secondary waves. Therefore,
influenced by plasma inhomogeneity, the secondary waves move
toward smaller $k$ for $L>0$ and toward larger $k$ for $L<0$. The
sign of a plasma gradient determines the spectrum of Langmuir
waves and the operating physical processes. However, independently
on the sign of the plasma gradient (figs.~\ref{fig5} and \ref{fig6})
the system approaches a steady state faster than in case of a
homogeneous plasma.

We consider the case of decreasing plasma density. One of the
conspicuous features of the turbulence evolution in inhomogeneous
plasma is the existence of quasi-steady states. The spectrum of
Langmuir waves remains almost constant at the time scale greater
than the time required for a decay. To obtain steady state, one
needs a constant source of Langmuir waves \cite{Sakharov79}. In
fig.~\ref{fig5} we see that during some time the values of
$E^{\pm}_{l}$ do not follow the typical oscillations associated
with a decay, as seen for instance in fig.~\ref{fig1}. In
homogeneous plasma these values undertake oscillations with a
period defined by the intensity of waves. The formation of the
quasi-steady state is explained by the following observations: Due
to three wave interaction, Langmuir waves spread over $k$ space,
with each scattering  resulting in lower intensities for a given
$k$. The rate of decay is essentially determined by the intensity
of the Langmuir waves and generally decreases with each  decay. At
some instant, the rate of decay becomes comparable with the effect
induced by the plasma inhomogeneity. From this time,
$t\approx 0.05$ s, the decay is suppressed by the shift due to the plasma
density gradient. With time, this equilibrium is  broken at
$t\approx 0.09$ s, and the oscillations proceed. This quasi-steady state
is also clearly seen in the spectrum of ion-sound waves. The
ion-sound waves corresponding to the decay of back-scattered
Langmuir waves appear only after breaking of the equilibrium. It
should be noted that the stabilization of the decay instability by
the plasma inhomogeneity has not been considered before. The
stabilization of a decay is somewhat analogous to stabilization of
scattering off ions, the only effective process for an isothermal
plasma. The stabilization of scattering off ions by a plasma
inhomogeneity has been considered previously \cite{Sakharov79}. In
a more general cases, when both nonlinear processes are allowed,
the compensation of a decay by a plasma inhomogeneity enables
scattering off ions to be seen at early stages of turbulence. The
quasi steady state spectrum of Langmuir turbulence slowly changes
due to scattering off ions.

The opposite case, with positive sign of plasma gradient, $L>0$,
leads to Landau damping of back-scattered waves on thermal
electrons with negative velocity. Indeed, as a result of
absorption, accelerated electrons are seen in the range
$-8\,v_{Te}<v<-4\,v_{Te}$ (fig. \ref{fig6}). Further
development of wave turbulence is suppressed and a quasi-steady
state is not formed. Similar to the large plasma gradient case,
with $L>0$,  the significant part of primary Langmuir waves are
re-absorbed by the beam. The maximum velocity of the plateau, see
fig.~\ref{fig6}, is clearly larger than in the case of a
homogeneous plasma.

\subsection{Weak plasma plasma inhomogeneity}

\begin{figure}
\begin{center}
\includegraphics[width=70mm]{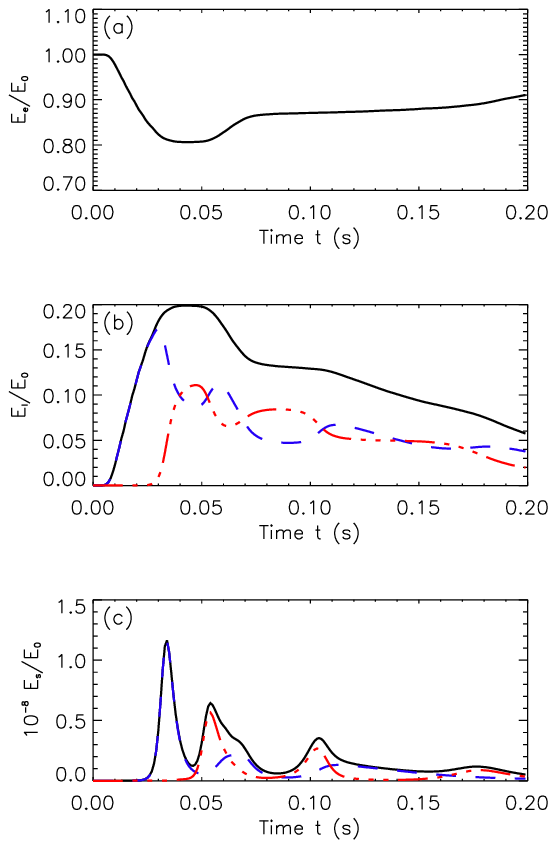}
\includegraphics[width=70mm]{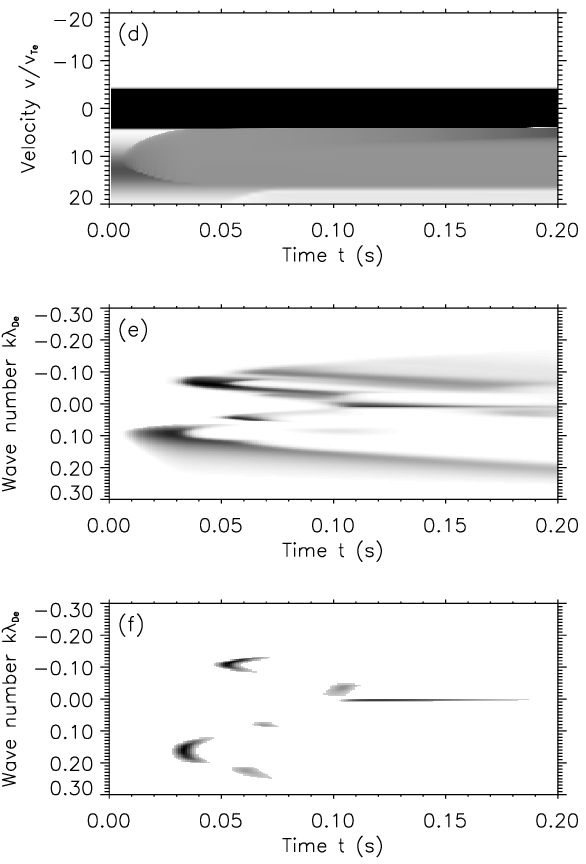}
\end{center}
 \caption{The same as fig.~\ref{fig1} but for a decreasing plasma density.
Weak plasma inhomogeneity $L=-1\times 10^9$ cm.}
 \label{fig7}
\end{figure}

\begin{figure}
\begin{center}
\includegraphics[width=70mm]{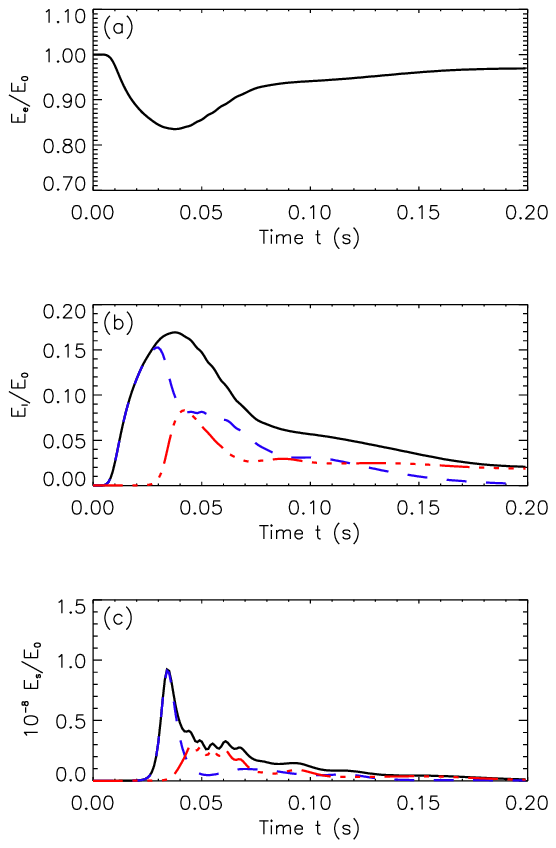}
\includegraphics[width=70mm]{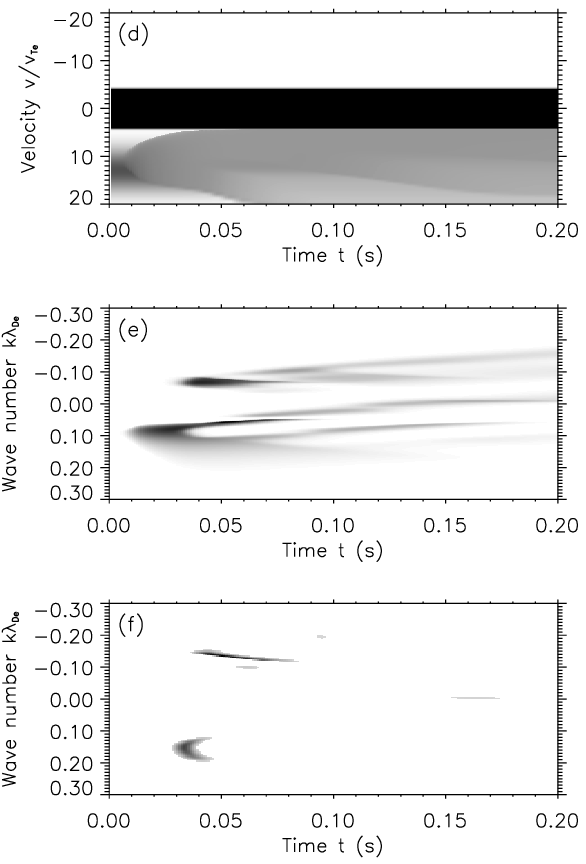}
\end{center}
 \caption{The same as fig.~\ref{fig1} but for an increasing plasma density.
Weak plasma inhomogeneity $L=1\times 10^9$ cm.}
 \label{fig8}
\end{figure}

When  the plasma inhomogeneity is decreased, the overall pictures tends to
be more and more similar to the case of homogeneous plasma.
However, there are some features, which makes the spectrum
different from homogeneous plasma (figs.~\ref{fig7} and
\ref{fig8}). The weak plasma inhomogeneity effectively governs the
turbulence in the range of small $k$.

The negative plasma gradient (fig.~\ref{fig7}) again leads to some
quasi-steady states, but for $L=-1\times 10^9$ cm it requires a
few acts of decay to reach compensation of three wave decay by
plasma inhomogeneity. The noticeable part of primary Langmuir wave
energy goes directly to background plasma. The other waves
experience decay and are then absorbed by the beam. We also see an
appearance of accelerated electrons as in the case of increasing
plasma density (fig.~\ref{fig7}). However, the physics behind this
process is different. Both primary Langmuir waves and back-scattered
secondary waves are now absorbed by the beam. Thus, a relatively weak
plasma inhomogeneity, with $L=1\times 10^9$, cm prevents further
decay of Langmuir waves.

The notable point is  the influence of a plasma inhomogeneity in the
region of small $k$, where decay is prohibited but scattering off
ions continues building of Langmuir wave condensate. This high
level of plasma waves accumulated near $k\approx 0$  is shifted by
the plasma gradient out of the region where decay prohibited. As a
result, Langmuir waves  may again decay, producing ion-sound waves.
This interplay of nonlinear processes leads to generation of sound
waves with very small $k$. Since the damping rate is smaller for
lower $k$, the intensity of these waves can be quite high
(figs.~\ref{fig7} and \ref{fig8}).

The plasma inhomogeneity also affects the spectral distribution of
ion-sound waves. Comparing cases with homogeneous and
inhomogeneous plasmas, we conclude that a plasma inhomogeneity
increases the efficiency of ion-sound wave generation. The other
interesting feature is the appearance of small wavenumber
ion-sound waves in case of weak plasma gradients. The damping rate
$\gamma^s_k$ decreases with $k$, implying that long wavelengths
are most likely to be observed late in the evolution of the
turbulence.  The distribution of these waves is a result of both
nonlinear processes (decay and scattering) acting differently at
low $k$.

\section{Discussion and main results}\label{sec:discuss}

As we see, due to a plasma gradient the Langmuir wave spectrum is
shifted in $k$ space with different processes being active. Thus,
a plasma inhomogeneity is effective in switching between different
processes in the system. Indeed, for sufficiently small $L$
(strong gradient) the plasma inhomogeneity prohibits any nonlinear
processes. Instead, the electron-Langmuir wave interaction becomes
important. For $L>0$ the Langmuir waves are absorbed back by the
beam, while for $L<0$ Langmuir waves are absorbed by thermal
electrons via Landau damping. Increasing $|L|$ (reducing the
gradient) we activated the next process, the three wave decay of a
Langmuir waves. Depending on plasma gradient, the Langmuir
turbulence makes a fixed number of oscillations. The plasma
inhomogeneity controls the rate of scattering off ions and decay
at small  $k$, where decay and scattering becomes equally
important.

Our calculations demonstrate that the  Langmuir wave spectrum in general
reaches a
quasi-stationary state more rapidly in an inhomogeneous plasma that in a
homogeneous one.

The positive plasma gradient case shows that the amount of energy
that is released in form of Langmuir waves is much smaller than
that in case of a homogeneous plasma. The positive plasma gradient
also leads to appearance of accelerated electrons in the plasma
\cite{Ryutov69}.

The ion-sound waves are heavily damped in plasmas with comparable
ion and electron temperatures. First, this leads to a low level of
ion-sound waves in comparison to that of Langmuir waves. Indeed, the
energy accumulated in ion-sound waves is a small fraction of
initial beam energy. Second, the ion-sound waves are seen as
bursts with small time duration. Obviously, the plasma
inhomogeneity can also influence the ion-sound turbulence. The
interesting observation is that the period of time the bursts of
ion-sound waves exist is dependent on plasma gradient. Generally,
the bursts of ion-sound waves are longer in time in inhomogeneous
plasma than in homogeneous. Therefore Langmuir wave turbulence
loses energy via ion-sound waves more intensively in comparison
with the homogeneous plasma case. It may be that anomalous spectra of
ion sound waves observed by incoherent radar scattering can be
explained in terms of decay from Langmuir waves \cite{forme}, and
we anticipate that naturally occurring plasma density gradients in
the upper parts of the ionosphere can contribute also to these
processes.

In view of application to astrophysical plasma, the plasma
turbulence is normally observed via non-thermal radio emission
\cite{sol_phys,hoang,thejappa}. The efficient emission mechanism
giving radio emission at double plasma frequency is the
coalescence of two Langmuir waves. In order to obtain high
intensity of radio emission one has to supply high level of waves
propagating at an angle to primary Langmuir waves. In this view
low wavenumber Langmuir waves are the most effective. As shown, a
plasma inhomogeneity successfully governs the evolution of the 
small wavenumber plasma
waves, and provides us with a better understanding also of the Langmuir
turbulence responsible for radio emission.

\section{Conclusions}\label{sec:concl}

We demonstrated that a plasma inhomogeneity can act as a control
parameter and play a crucial role in the development of weak
Langmuir turbulence, by triggering various processes that affects
the turbulence. The plasma inhomogeneity generally leads to a
decline of turbulence, therefore a steady state of weak turbulence
is reached faster than in case of homogeneous plasma. Selective
effects of Landau damping of Langmuir waves, or self acceleration
of electrons in a beam can easily be activated by proper choice of
a density gradient.

The plasma inhomogeneity is the main 
process that is capable of switching between the decay and
scattering off ions at very low wavenumbers. This part of the
Langmuir wave spectrum is important for radio emission, and a
plasma inhomogeneity can therefore be significant in adjusting the
flux of radio emission from the plasma. Generally, we expect that
the radio emission should be stronger for inhomogeneous plasmas.
It might be appropriate to mention that the relative importance of electron
temperature gradients is negligible in comparison to density gradients with
the same length scale $L$. When considering the last term in
(\ref{wave_kin}) we have an effective force acting on a wave packet being
$\partial \omega_k/\partial x \sim \omega_{pe}/L$ for the density
gradient, while it is $\partial \omega_k/\partial x \sim
\omega_{pe}(k\lambda_{De})^2/L$ for the homogeneous density with
inhomogeneous background electron plasma temperature. 
For this term in (\ref{wave_kin}), a temperature
gradient gives a negligible contribution as long as $(k\lambda_{De})^2\ll
1$. The constraint implied in the WKB-approximation (\ref{WKB}) must be
fulfilled also, but this is  a rather weak requirement, considering the
large values of $L$ used in the present analysis. We must, however,  keep in
mind that both $St_{decay}$ and  $St_{ion}$ are  temperature
dependent, and that temperature gradients can affect these terms. 

The presence of inhomogeneity affects the decay process, and as a
result ion-sound turbulence is generated more effectively. In
particular, the plasma inhomogeneity can give rise to a
quasi-steady state of decay interaction, when the shift of the
Langmuir wave spectrum due to inhomogeneity suppresses further
developments of decay. From an application point of view, it is
interesting that in our case a weak plasma inhomogeneity
stimulates generation of ion sound waves
with very low wavenumbers. We took particular care to
consider physically realistic parameter values, in the present
case some relevant for the solar corona. We find it of particular interest
in this context, that the time scale for the initial 
evolution of the waves and the electron beam which is driving the process 
is short, of the order of 0.02 -- 0.05 s, i.e.\
comparable to a typical electron--ion collision time, 
$\nu_{ei}^{-1}\geq 1/50$~s. A collisionless plasma model for this type 
of processes is therefore justified.

\begin{acknowledgments}
The work was in part supported by the Research Council of Norway
(NFR). We also received support from the Programme for Supercomputing,
through a grant for computing time.
\end{acknowledgments}

\end{document}